\documentclass[twocolumn]{jpsj2}
\def\rfo{$R$Fe$_2$O$_4$}
\def\lufo{LuFe$_2$O$_4$}
\def\f3o4{Fe$_3$O$_4$ }
\def\fetwo{Fe$^{2+}$  }
\def\fethr{Fe$^{3+}$  }
\def\PRB{Phys. \ Rev. \ B  }
\def\PRL{Phys. \ Rev. \ Lett.  }
\def\JPSJ{Jour. \ Phys. \ Soc. \ Jpn. }

\title{Electronic Ferroelectricity and Frustration}

\author{
Sumio \textsc{Ishihara}\thanks{Corresponding author. E-mail: ishihara@cmpt.phys.tohoku.ac.jp} }

\inst{
Department of Physics, Tohoku University, Sendai 980-8578 Japan 
}

\abst{
Beyond a conventional classification of ferroelectricity, there is a class of materials where electronic degrees of freedom and electronic interactions are directly responsible for electric polarization and ferroelectric transition. This is termed electronic ferroelectricity. In this article, we review electronic ferroelectricity from a view point of frustration. 
Experimental and theoretical examinations in spin driven ferroelectric materials, recently termed multiferroics, are introduced.  
Spin frustration caused by competing magnetic interactions is of prime importance for this type of ferroelectricity. 
Charge driven ferroelectricity where electronic charge order induces electric polarization is reviewed. 
In particular, exotic dielectric and magneto-dielectric properties in layered iron oxides are focused on. 
Through a number of recent experimental and theoretical researches, charge fluctuation due to frustrated geometry plays essential roles on electronic ferroelectricity in this compound. 
}

\kword{Ferroelectricity, Multiferroics, Frustration, Charge Order}

\begin{document}
\maketitle

\section{Introduction} 

\noindent

We have learned from solid-state textbooks that ferroelectricity is defined by appearance of macroscopic electric polarization and its reversibility by applying external electric field. Since the discovery of ferroelectricity in Rochelle salts in 1920~\cite{valasek20}, a great number of ferroelectric compounds have been found and investigated intensively from view points of fundamental physics as well as technological applications. 
In terms of mechanisms of the phase transition, ferroelectric compounds are formally classified into the following groups; 
1) displacive-type ferroelectricity where relative displacement of negative and positive ions 
is responsible for a macroscopic electric polarization, 
%
and 2) order-disorder type ferroelectricity where the phase transition occurs by cooperative alignment of permanent dipole moments. 
Hydrogen-bond type ferroelectricity is often regarded as an order-disorder type, although its mechanism is still under debate.

For a long time, the Landau-Devonsher type phenomenological theory and the lattice dynamics calculation based on the shell model have revealed some aspects in structural and ferroelectric transitions. 
In 1990s, theoretical studies in ferroelectricity have been improved revolutionary; the first-principle band structure calculations revealed considerable contribution of electronic degree of freedom to ferroelectric transition~\cite{cohen90,cohen92}. 
In particular, through the calculations of electric polarization and the Born effective charge based on the Berry phase theory, electronic hybridization between transition-metal and oxygen ions considerably supports the displacive-type ferroelectric transition~\cite{resta92,kingsmith93,vanderbilt93,resta94}.  

Beyond the standard classification of ferroelectrics, it is known that, in some types of ferroelectric materials, electron degrees of freedom and/or electronic interactions directly give rise to a macroscopic electric polarization and ferroelectric transition. 
Here, we call this phenomenon electronic ferroelectricity and these materials electronic ferroelectric compounds.  
There are the two types of the electronic ferroelectricity;  
1) spin driven ferroelectricity where magnetic interaction and magnetic ordering mainly cause the ferroelectric transition. 
This class of materials belongs to a Mott insulator with an integer number of electrons per site. 
Since magnetic order parameter is a primary order parameter, this is an improper ferroelectrics.  
The discovery of ferroelectricity in TbMnO$_3$ in 2003~\cite{kimura03} opened a recent 
boom of this type of ferroelectricity where observed cross correlation between ferroelectricity and magnetism is often termed multiferroics. 
It is recognized that spin frustration is of primly importance for this type of ferroelectricity. 
2) Charge driven ferroelectricity where electronic charge degree and charge order are responsible for an electric polarization. Some materials in this type belong to the so-called quarter-filled system where an average electron number per site is a half integer.  
This ferroelectricity has been observed in some transition-metal compounds and low-dimensional charge-transfer organic salts. 
Since the electric polarization in most of this class of materials is associated with charge ordering, this ferroelectricity is caused by electron correlation and/or electron-lattice interactions. 
It has been thought since long time ago that magnetite Fe$_3$O$_4$ belongs to this type of ferroelectricity; anomalous dielectric behaviors observed near the Verwey transition, i.e. the charge ordering transition of Fe $3d$ electrons, have attracted long-time attention. 
A two dimensional version of the Verwey transition is seen in the layered iron oxide \lufo~\cite{ikedanature} which is also regarded as an electronic ferroelectric material. 

In this paper, recent progress of electronic ferroelectricity in correlated electron system is reviewed. In particular, theory and experiment are surveyed from view point of geometrical frustration.  Instead of well-studied electronic contributions in standard ferroelectric materials, such as hybridization effects between cations and anions, and effects of lone-pair electrons in displacive type ferroelectricity, we focus on the electronic effects where electron degrees of freedom are concerned directly in ferroelectricity. We restrict ourselves to reviewing materials and phenomena where spin and charge degrees of freedom in electrons are responsible for ferroelectricity. 

In Sect.~2, we provide a brief overview of theory and experiment in spin driven ferroelectricity. 
Charge driven ferroelectricity are introduced 
in Sect.~3. 
One of the main topics in this article is to introduce a prototypical example for the charge driven ferroelectricity, layered iron oxide. 
Recent experimental and theoretical studies in this series of materials are surveyed in Sect.~4. 
Section 5 is devoted to the concluding remarks and future issues. 
Also see other reviews for multiferroics and charge driven ferroelectricity in Refs.~\cite{cheong07,brink08}

\section{Spin driven ferroelectricity}

It is generally recognized that most of the ferroelectricity are realized in band insulators where spin degree of freedom of electrons is quenched. 
In some Mott insulating magnets, 
magnetoelectric effect was known to be observed~\cite{dzyaloshinskii59,rado61,date61,mitsek63}. 
Coexistence of ferroelectricity and magnetism was also recognized long time ago and is termed magneto-ferroelectricity~\cite{smolenskii82,hill00}. 
A number of magneto-ferroelectric compounds were discovered and have been studied extensively, in particular, perovskite oxides such as BiFeO$_3$~\cite{bifeo}, hexagonal manganites $R$MnO$_3$~\cite{ymno}, rare-earth manganites $R$Mn$_2$O$_5$~\cite{kohn1,kohn2}, boracites M$_3$B$_7$O$_{13}$X~\cite{boracite}, and others. 
In some of the magneto-ferroelectric compounds, ferroelectric transition temperature is far above magnetic transition temperature. 
Although magneto-electric effects arise below the magnetic transition temperature, it is believed that two transitions occur almost independently. 
Ferroelectricity discovered in orthorhombic perovskite manganite TbMnO$_3$~\cite{kimura03} has triggered recent extensive studies in coexistence of ferroelectricity and magnetism termed multiferroics. 
We introduce electronic structures and dielectric/magnetic properties in manganites as a prototypical example of spin-driven ferroelectricity with geometrical frustration. 

\begin{figure}[tb]
\begin{center}
\includegraphics[width=0.9\columnwidth,clip]{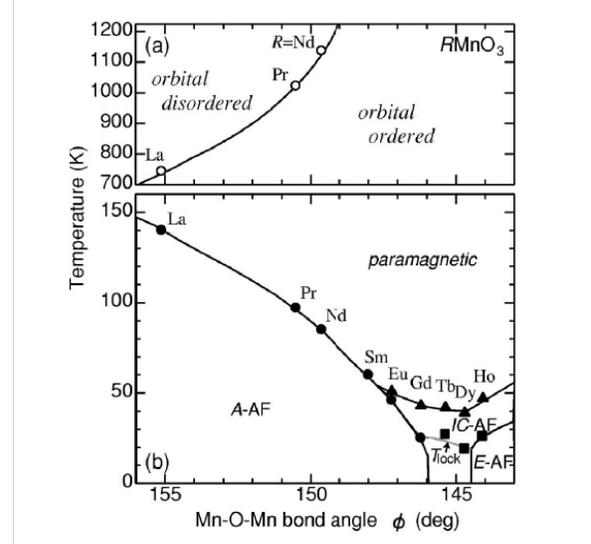}
\end{center}
\caption{ 
Orbital (a) and spin (b) ordering temperature of $R$MnO$_3$ 
as a function of the in-plane Mn-O-Mn bond angle~\cite{kimura03b}. 
A phase below $T_{\rm lock}$ between $144.5<\phi<146$ in (b) is the ferroelectric phase. 
}  
\label{fig:mnphase}
\end{figure}
Manganite with orthorhombic perovskite structure, $R$MnO$_3$ ($R$: rear-earth ion), is known as a parent compound for the colossal magneto-resistant manganite where a gigantic change in the electrical resistivity by applying magnetic field was found. Under a crystalline electric field in distorted perovskite structure, an electron configuration for the 3$d$ orbitals in a Mn ion is $(t_{2g})^3 (e_g)^1$, 
and a system is a Mott insulator. 
A systematic change in crystal structure by changing $R$ ion is characterized by the Mn-O-Mn bond angle; 
a deviation of the bond angle from 180$^\circ$ becomes steep by changing $R$ from La to Ho in the periodic table. 
The spin and orbital phase diagrams as a function of the Mn-O-Mn bond angle 
are presented in Fig.~\ref{fig:mnphase}. 
In LaMnO$_3$ with the largest bond angle in a series of $R$MnO$_3$, 
the staggered order of the $d_{3x^2-r^2}$ and $d_{3y^2-r^2}$ orbitals in the $ab$ plane occurs at $T_{OO}=750$K associated with the cooperative Jahn-Teller type lattice distortion. 
The A-type antiferromagnetic order, where ferromagnetic alignment in the $ab$ plane and antiferromagnetic one along the $c$ axis, appears below 145K. 
On the other side, in HoMnO$_3$ with the small Mn-O-Mn bond angle, the E-type antiferromagnetic order appears; the "up-up-down-down"-type four-fold spin order along the $b$ axis. 
Between LaMnO$_3$ and HoMnO$_3$, the ferroelectric polarization associated with a non-collinear spin structure exhibits. 
One representative material is TbMnO$_3$ where, with decreasing temperature, an  incommensurate magnetic order occurs at around 42K and the ferroelectric transition sets in at 28K where the magnetic structure is changed from collinear to noncollinear. 

\begin{figure}[tb]
\begin{center}
\includegraphics[width=\columnwidth,clip]{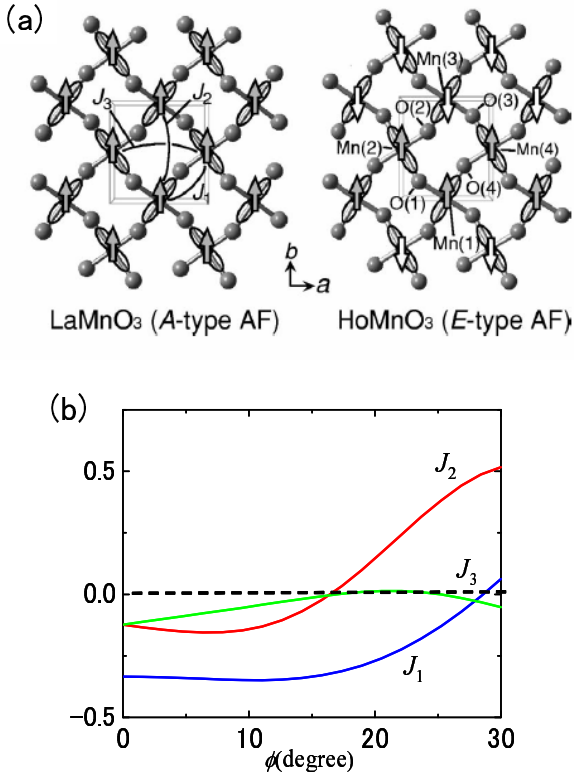}
\end{center}
\caption{(a) Schematic spin and orbital structures in LaMnO$_3$ and HoMnO$_3$, 
and (b) bond angle dependence of the superexchange interactions 
in the $ab$ plane calculated by the spin-orbital model~\cite{kimura03b}.
A horizontal axis represents deviation of the Mn-O-Mn bond angle from 180 degree. 
Positive and negative values of the exchange interactions correspond to antiferromagnetic and ferromagnetic interactions, respectively.}
\label{fig:rmno}
\end{figure}
Spin and orbital structures in $R$MnO$_3$ are shown in Fig.~\ref{fig:rmno}~\cite{kimura03b}. 
Due to the staggered orbital order, the superexchange interaction between the nearest neighbor (NN) Mn ions, $J_1(<0)$, is ferromagnetic. 
An antiferromagnetic superexchange interaction between the second NN Mn ions along the $a+b$ axis, $J_2(>0)$, is also caused through the Mn-O-O-Mn exchange path. 
In Fig.~\ref{fig:rmno}(b), theoretically calculated exchange interactions as functions of the Mn-O-Mn angle are presented. Here, the exchange interactions are evaluated by the perturbational calculation with respect to the electron hopping between the NN Mn and O sites and the long-range orbital order is assumed. 
It is shown that with decreasing the Mn-O-Mn bond angle, $|J_1|$ decreases and $|J_2|$ increases due to a large overlap integral between the two O $2p$ orbitals along the $a+b$ axis. 
Therefore, this system is regarded as a frustrated spin system described by 
the so-called $J_1-J_2$ localized spin model. 
Ferromagnetic (up-up-down-down type magnetic) structure corresponding to the A-type (E-type) antiferromagnetic order, and collinear and non-collinear spin structures with long periodicity between them are naturally understood in a view point of frustrated spin system where magnitude of the frustration is controlled by the Mn-O-Mn bond angle corresponding to $|J_2/J_1|$.   
After discovery of the spin driven ferroelectricity in TbMnO$_3$, ferroelectricity and multiferroelectricity were examined and reexamined in several transition-metal compounds with non-collinear magnetic orders, 
such as, $R$Mn$_2$O$_5$~\cite{kobayashi03,hur04}, Ni$_3$V$_2$O$_8$~\cite{lawes05}, CuFeO$_2$~\cite{kimura06,seki07}, RbFeMoO$_4$~\cite{kenzelmann07}, CuO~\cite{kimura08} and others. 

\begin{figure}[tb]
\begin{center}
\includegraphics[width=0.8\columnwidth,clip]{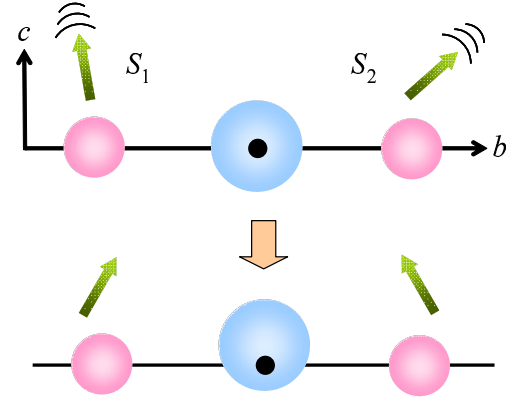}
\end{center}
\caption{ 
Schematic view of the inverse Dzyaloshinskii-Moriya interaction mechanism. A paramagnetic and paraelectric phase (above) and a non-collinear magnetic ordered ferroelectric phase (below).
Blue large and pink small circles represent the oxygen ions and the transition-metal ions, respectively. 
}  
\label{fig:dm}
\end{figure}

As origin of the ferroelectricity in such frustrated spin systems, 
the following two scenarios are proposed: 
1) the inverse Dzyaloshinskii-Moriya (DM) interaction mechanism or the spin current mechanism~\cite{katsura05, sergienko06, mostovoy06} , and 2) the exchange striction mechanism~\cite{arima06,aliouane06, sergienko06b}. 
In both cases, an electric polarization appears to gain the magnetic exchange energy.  
A phenomenological theory was given in Ref.~\cite{harris07}. 

Let us introduce the inverse DM mechanism.  
Consider a cycloidal spin structure where spins rotate in the $bc$ plane 
along the $b$ axis, and focus on two NN spins along the $b$ axis, termed $\vec S_1$ and $\vec S_2$ [see Fig.~\ref{fig:dm}]. 
The DM interaction Hamiltonian between the two spins are written as~\cite{dzyloshinskii58, moriya60}
\begin{equation}
{\cal H}_{DM}=\vec D \cdot \left ( \vec S_1 \times \vec S_2 \right ),  
\label{eq:dm}
\end{equation}
where $\vec D$ is the DM vector which is of the order of $(\lambda t_{pd}^4/\Delta E^4)$ with the relativistic spin-orbit coupling $\lambda$, the electron transfer integral between NN O $2p$ and the transition-metal $3d$ orbitals, $t_{pd}$, and the energy separation between the ground and excited states, $\Delta E$. 
Above the spin ordering temperature, it is assumed that there is a reflection symmetry at a center of the bond connecting sites 1 and 2, and the DM vector is zero, $\vec D=0$. 
With decreasing temperature, a cycloidal spin order sets in at a certain temperature due to frustrated exchange interactions, and a vector chirality $\vec S_1 \times \vec S_2$ becomes finite. 
Then in order to gain the DM interaction energy in Eq.~(\ref{eq:dm}), a spontaneous breaking of inversion symmetry along the $c$ axis is induced and $\vec D$ becomes finite. 
This is the inverse process of the usual DM interaction in magnets. 
There, the DM vector is finite, $\vec D \ne 0$, due to the low symmetry crystal structure in all temperature range. 
Above the magnetic transition temperature, the vector chirality is zero, $\vec S_1 \times \vec S_2=0$, and below the temperature, a non-collinear spin structure with $\vec S_1 \times \vec S_2 \ne 0$ is induced in order to gain the DM interaction energy. 
%
%
In this scenario, a direction of the polarization $\vec P$ is given by~\cite{katsura05, sergienko06, mostovoy06} 
\begin{equation}
\vec P \propto \vec d_{12} \times \left ( \vec S_1 \times \vec S_2 \right ) , 
\label{eq:dmp}
\end{equation}
where $\vec d_{12} $ is a vector connecting $\vec S_1$ and $\vec S_2$. 
In the cycloidal spin structure introduced above, a macroscopic polarization appears along the $c$ axis. 
The spin current mechanism is the electronic version of this mechanism; the electronic cloud breaks the inversion symmetry in the limit of heavy ion mass. 
It is noted that the relation between an electric dipole moment and vector chirality was first given 
in a theoretical study of optical absorption by magnetic crystal in Ref.~\cite{moriya68}. 
A magnetoelectric effect based on the DM interaction was examined in ZnCr$_2$S$_4$ where the observed magnetoelectric response is shown to depend on a direction of the spiral plane~\cite{shiratori80,shiratori80b}. 

\begin{figure}[tb]
\begin{center}
\includegraphics[width=0.9\columnwidth,clip]{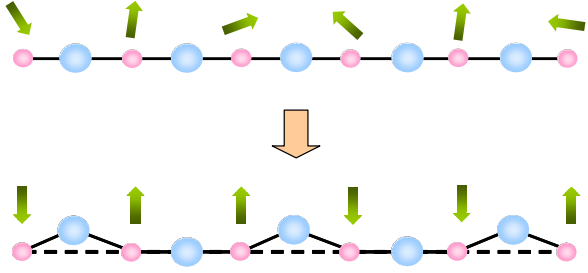}
\end{center}
\caption{ 
Schematic view of the magneto-striction mechanism for the up-up-down-down collinear spin ordered phase. Blue large and pink small circles represent the oxygen ions and the transition-metal ions, respectively.}
\label{fig:dm2}
\end{figure}
As for the exchange striction mechanism, the coupling between the electric polarization and spins are caused by the symmetric exchange interaction $J \sum_{\langle ij \rangle} \vec S_i \cdot \vec S_j $ 
where the exchange constant $J$ is of the order of $(t_{pd}^4/\Delta E^3)$.  
Consider a chain where the transition-metal ions M and oxygen ions O are aligned alternately, and the exchange interaction between the NN M ions caused by the superexchange processes through the O ion in between the two M ions. 
A commensurate collinear magnetic order such as the up-up-down-down structure, as shown in Fig.~\ref{fig:dm2} is considered. 
Favorable bond angles in the M-O-M bonds depend on spin configurations in M ions. 
Here we assume that a large (small) bond angle is favored in the ferromagnetic (antiferromagnetic) spin configuration to gain the exchange energy. This is realized in the exchange interaction with the orbital degree of freedom in a M ion. 
As a result, the O ions between the antiparallel spins are distorted spontaneously, and the inversion symmetry in the system is broken. 

The relation between the spin structure and the electric polarization was examined by the neutron diffraction experiments. 
Temperature dependence of the magnetic structure in TbMnO$_3$ was determined in Ref.~\cite{kenzelmann05}; a collinear longitudinal incommensurate spin order is changed into a transverse incommensurate spiral order in the $bc$ plane at the ferroelectric transition temperature. 
A more direct relation between non-collinear spin structure and electric polarization is examined in Ref.~\cite{yamasaki07}. By using the polarized neutron diffraction method into TbMnO$_3$, it is revealed that spin chirality $\vec S_1 \times \vec S_2$ changes its sign by flipping the electric polarization by applying external electric field. 
These experimental results suggest that the electric polarization in TbMnO$_3$ is driven by the inverse DM mechanism. 

Other mechanisms of coupling between electric polarization and spin order have been proposed. 
The spin-polarization couplings in the case where the orbital magnetic moment is not quenched, such as a $(t_{2g})^n$ configuration, are examined in Refs.~\cite{onoda07} and \cite{arima07}. 
Considered exchange couplings between the NN transition-metal ions are 
of the order of $\lambda t_{pd}^3/\Delta E^3$ and $\lambda^2 t_{pd}^2 /\Delta E^3$, respectively. 
The latter mechanism is applied to the multiferroic CuFeAl$_4$. 

Finally, we review the magnetic and dielectric phase diagram in $R$MnO$_3$ presented in Fig.~\ref{fig:mnphase}(b) from the view point of frustration. 
As mentioned above, systematic magnetic structure change by changing $R$ ions is naturally understood by the frustrated $J_1-J_2$ model; the A-type AFM and E-type AFM orders are resulted mainly from the NN ferromagnetic exchange interaction $J_1$ and the NNN AFM interaction $J_2$, respectively. Between the two phases, the two interactions are competed with each other, and a non-collinear spin structure where $\vec S_i \times \vec S_2 \ne 0$ is realized. By inducing a finite value of the DM vector, $\vec D \ne 0$, the energy reduction due to the DM interaction is brought about. In the case where a uniform component of the $D$ vector remains in a whole crystal, a macroscopic electric polarization appears. 
In this sense, it is concluded that spin frustration is an origin of the spin driven ferroelectricity. 

\section{Charge driven ferroelectricity}

Contrary to the spin driven ferroelectricity, not so many experimental and theoretical papers have been published for the charge driven ferroelectricity so far. 
This is because most of this type of ferroelectric materials belongs to the quarter- or fractional- electron filled charge order system.
Since charge carriers easily fluctuate around localized sites, this ferroelectricity is fragile and dielectric properties are diffusive. 
In general, it is difficult to measure the hysteresis curve by applying electric field due to leak current. 
Some transition-metal oxides~\cite{brink08} and organic salts~\cite{horiuchi08} are thought to be this type of ferroelectric material because of their structural and dielectric anomalies. 
Perovskite manganites Pr$_{1-x}$Ca$_{x}$MnO$_3$ ($x \sim 0.5$)~\cite{feremov04, lopes08}, layered perovskite manganites Pr(Sr$_x$Ca$_{1-x}$)$_2$Mn$_2$O$_7$ ($x \sim 0.1$)~\cite{tokunaga06}, and quasi-one dimensional organic salts (TMTTF)$_2$$X$ ($X=$PF$_6$, AsF$_6$)~\cite{javadi98,monceau01} are plausible candidates. 
From the view point of the geometrical frustration, we briefly introduce in this section magnetite Fe$_3$O$_4$ and quasi-two dimensional organic salt $\alpha$-(BEDT-TTF)$_2$I$_3$. 
Anomalous dielectric and magneto-dielectric properties in layered iron oxides $R$Fe$_2$O$_4$ with frustrated geometry are introduced in the next section. 

Magnetite \f3o4\ is one of the most famous insulating magnet. 
Chemical formula is often written as Fe$^{3+}$(Fe$^{2+}$Fe$^{3+}$)O$_4$ 
where equal amount of \fetwo and \fethr occupy the so-called B sites in the spinel crystal structure with frustrated geometry. 
The well known phenomenon in this compound is the Verwey transition~\cite{verwey39}; 
a metal-insulator transition at around 120K which is believed to be the charge order transition of \fetwo and \fethr in the spinel B sites. 
An important role of the geometrical frustration in the Verwey transition and a macroscopic number of degenerate charge ordered states were first suggested by Anderson~\cite{anderson56}, and a condition which should be satisfied in the charge order pattern is termed the Anderson's condition. 

Anomalous dielectric properties were reported since 1970's. 
Magnetoelectric responses measured below the Verwey transition were interpreted by a model with spontaneous electric polarization along the $b$ axis~\cite{rado75,rado77,miyamoto88,miyamoto93}.
Ferroelectric hysteresis loop along the $c$ axis was also observed in Ref.~\cite{kato82}.
Instead of such experimental reports about ferroelectricity, 
detailed ferroelectric properties and its mechanism are not still unveiled. 
Since 2000, the Veywey transition and the associated charge order have been reinvestigated by the recently developed experimental techniques. 
The resonant x-ray scattering experiments reported no evidence of charge disproportionation 
within an experimental time scale~\cite{garcia00,garcia04}. 
Combined powder neutron and x-ray diffraction study suggest an evidence of the charge order where Fe valences are 2.4+ and 2.6+, and its ordering patter does not satisfy 
the Anderson's condition~\cite{wright01}. 
As for theoretical examinations in terms of ferroelectricity, 
a possible mechanism of ferroelectricity is considered from view point of site-centered and bond-centered charge orders in Ref.~\cite{brink08}. 
A first-principle band structure calculation proposed that the ferroelectricity below the Verwey transition temperature is induced by the noncentrosymmetric charge order~\cite{yamauchi09}. 
In spite of these experimental and theoretical investigations introduced above, origin of the ferroelectricity and a detailed charge order pattern are still under debate.  

\begin{figure}[tb]
\begin{center}
\includegraphics[width=\columnwidth,clip]{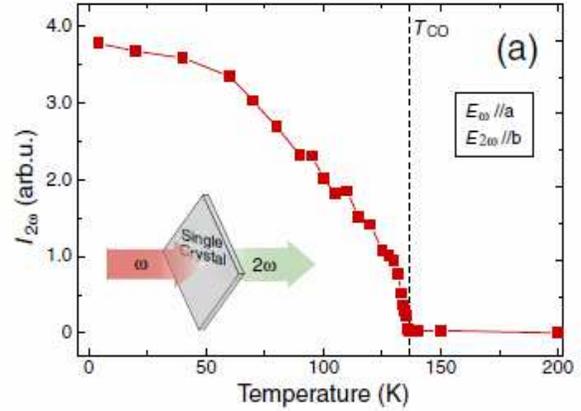}
\end{center}
\caption{ 
Temperature dependence of SHG intensity in $\alpha$-(BEDT-TTF)$_2$I$_3$~\cite{yamamoto08}.
}
\label{fig:yamamoto}
\end{figure}
Another example of the electronic ferroelectricity is seen in the quasi-two dimensional organic salt $\alpha$-(BEDT-TTF)$_2$I$_3$. 
The crystal structure is a stacked insulating $I_3$ layers and (BEDT-TTF)$_2$ anisotropic triangle layers~\cite{dressel94}. 
Average charge carrier in a BEDT-TTF unit is 0.5 and the system corresponds to a quarter-filled quasi-two dimensional system. 
This material has much attention because of a metal-insulator transition at 135K. 
The low temperature insulating phase was identified as a charge ordered insulator confirmed by the NMR and Raman spectroscopy measurements~\cite{takano01,wojciechowski03}. 
The synchrotron x-ray diffraction experiment has revealed the so-called horizontal type charge order where four inequivalent BEDT-TTF units exist~\cite{kakiuchi07}. 
Several experimental and theoretical examinations for electronic structure and the charge ordered patterns have been done~\cite{iwai07,kino95,seo00,hotta03,miyashita08}. 

Recently, it is found experimentally that, in the low temperature charge ordered phase, the space inversion symmetry is broken. Yamamoto and coworkers reported that optical second-harmonic generation (SHG) signal appears below the charge ordering temperature~\cite{yamamoto08} as shown in Fig.~\ref{fig:yamamoto}. 
Thus, this material is a candidate of the charge driven electronic ferroelectricity, although the hysteresis curve is not obtained yet. 
Authors also demonstrate the time resolved pump-probe experiments of the SHG signal. 
The SHG signal is suppressed by photon-pumping and is recovered in the time scale of pico-second. 
This fast photo-response suggests that a main component of the electric polarization is not attributed to the lattice distortion but is due to the electronic charge order.

\section{Layered iron oxide}
\subsection{Dielectric and magnetic structures}

\begin{figure}[t]
\begin{center}
\includegraphics[width=\columnwidth,clip]{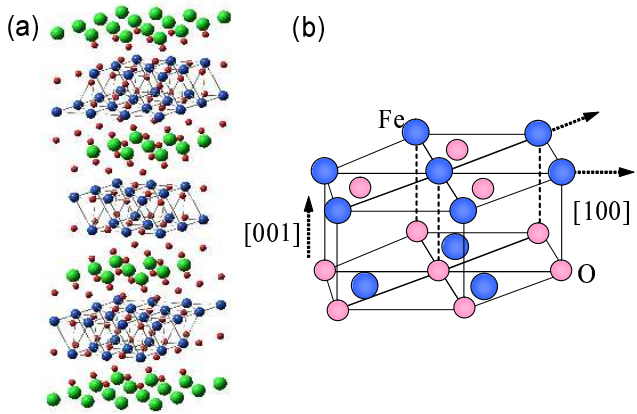}
\end{center}
\caption{ 
(a) Crystal structure of the layered iron oxide \rfo. 
Blue medium, red small and green large circles represent Fe ions, O ions and $R$ ions, respectively. 
(b) A pair of the Fe-O triangle layers termed the W-layer.
Blue large and pink small circles represent Fe and O ions, respectively. }
\label{fig:cryst}
\end{figure}
Layered iron oxide \rfo \ is one of the possible candidates for the charge driven ferroelectricity. 
This compound belongs to a homologous series 
($R$AO$_3$)$_n$$R$ABO$_4$ where A and B are the $3d$ transition-metal ions~\cite{kimizuka90}. 
This compound was first synthesized by Japanese and French groups~\cite{kato75,malaman75}, and has been studied intensively and extensively in their dielectric and magnetic properties. 
Crystal structure shown in Fig.~\ref{fig:cryst} consists of layered stacking of Fe-O double layers and $R$-O layers along the $c$ axis with the point group R$\bar 3$m. 
The dielectric and magnetic properties in \rfo \ are mainly responsible for Fe-O paired triangle layers, termed the W-layer, shown in Fig.~\ref{fig:cryst}(b). 
It is noted that in a W-layer, a Fe ion in the upper plane is not located above a Fe ion in the lower plane but above an O ion. 
A unit cell in high temperature phase includes three W-layers. 
In the chemical formula of \rfo, nominal Fe valence is $2.5+$, i.e. equal amount of \fetwo and \fethr in a triangular lattice. 
Therefore, this system is regarded as a frustrated charge-spin coupled system. 

\begin{figure}[t]
\begin{center}
\includegraphics[width=0.5\columnwidth,clip]{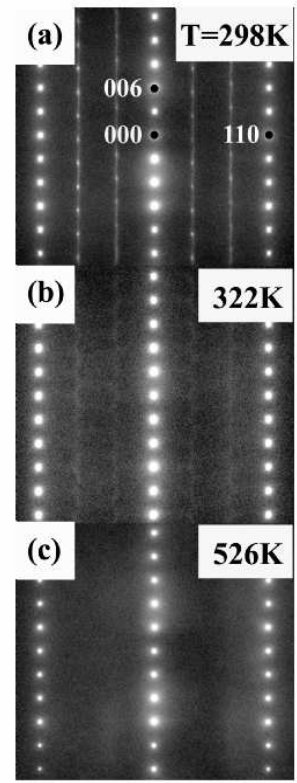}
\end{center}
\caption{ Temperature variation of electron diffraction patterns \lufo~\cite{matsuo08b,matsuo08}.
The beam incidence is almost parallel to the [1$\bar 1$0] direction.
The electron diffraction patterns are obtained at (a) 298K, (b) 322K and (c) 536K.}
\label{fig:eledif}
\end{figure}
Charge and lattice structures in \rfo \ has been studied by the electron diffraction, electron microscopy experiments and others~\cite{yamada97,yamada00,ikeda03,zhang07,zhang07b,murakami07}. 
In Fig.~\ref{fig:eledif}, electron diffraction pattern in \lufo \ are presented. 
With decreasing temperature, diffuse scattering appear along the $(1/3\ 1/3 \ l)$ below about 325K. 
Below about 300K, the super-lattice spots are observed at 
$(1/3 \ 1/3 \ 3m+1/2)$ inside of the diffuse streaks. 
These results imply the two-dimensional charge order between 325K and 300K and the three-dimensional long-range order with a three-fold periodicity in the plane below 300K. 
Pattern and periodicity of the charge order are sensitive to a rare-earth ion $R$. 
There is a sequential charge order transition in YFe$_2$O$_4$~\cite{ikeda03}; with decreasing temperature, the three-fold type charge order changes into other types of the charge order characterized by the electron diffraction peaks at $(0 \ 1/2 \ 0)$ and $(1/4 \ 1/4 \ 0)$. 
This result suggests a number of barely degenerate charge states and a competition between them. 
The three-fold charge order is also confirmed by the resonant x-ray scattering method at Fe K-edge~\cite{ikedanature}. 
A resonant diffraction intensity is observed at $(1/3 \ 1/3 \ 5.5)$ at 18K in \rfo. 
The scattering factor is well represented by a combination of the scattering factors in LuGaFeO$_4$ and LuCoFeO$_4$ where valences of Fe ion are 2+ and 3+, respectively, 
Therefore, it is claimed that valences of Fe ions are well separated and these are spatially ordered in the crystal lattice.

%
Magnetic structure has been examined by several experimental methods. 
In \lufo, the magnetization appears below about 250K~\cite{iida86,iida93,oka08,wu08,park09,wang07}. 
Detailed temperature dependence of the magnetic structure was studied by the neutron diffraction experiments~\cite{akimitsu79,funahashi84,shiratori92}. 
Around 270K in \lufo, two dimensional magnetic scattering along the $c^\ast$ axis was observed. 
Below 270K, the magnetic order becomes three-dimensional and the magnetic neutron diffraction peaks appear at $(1/3 \ 1/3 \ l)$ with integer $l$. 
The neutron diffraction pattern in \rfo \ is well analyzed by localized spin models with the three-dimensional ferrimagnetic order with Ising anisotropy. 
With further decreasing temperature, it is recently found that a peak intensity at (1/3 \ 1/3 \ 0) decreases around 170K~\cite{christianson08,kakurai}.

\begin{figure}[tb]
\begin{center}
\includegraphics[width=\columnwidth,clip]{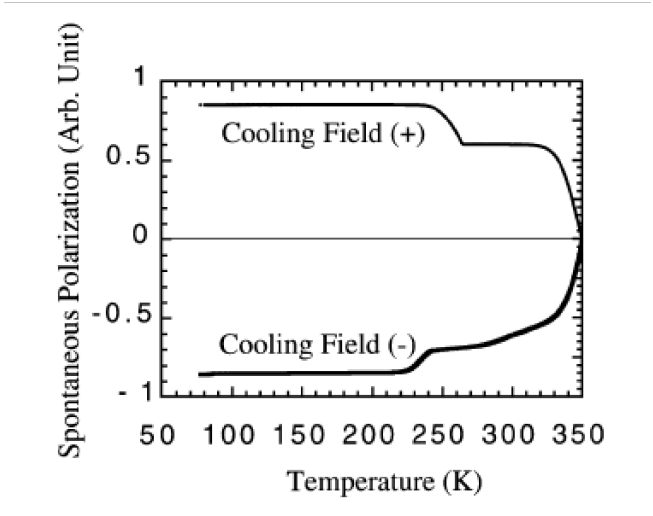}
\end{center}
\caption{ 
Temperature dependence of spontaneous electric polarization in\rfo estimated from pyroelectric current~\cite{ikeda00}.
Two curves represent the results after cooling in the electric field of the opposite directions. 
}
\label{fig:ppp}
\end{figure}
Electric polarization measured by the pyroelectric current is shown in Fig.~\ref{fig:ppp}~\cite{ikedanature,ikeda00}. 
Polarization is observed around the three-dimensional charge ordering temperature and further increases around 230K which is close to the three-dimensional ferrimagnetic ordering temperature. 
The results suggest that the electric polarization is responsible for the three-fold charge order, and there are strong coupling between the charge and magnetic degrees of freedom. 
Because of strong diffusive nature as mentioned latter, the $P-E$ hysteresis curve has not been obtained yet.
Recent neutron diffraction experiments claim that 
\lufo \ shows an antiferroelectric long-range order associated with ferroelectric short range order~\cite{angst08}. 
%
Strong magneto-dielectric coupling is also observed by measurements of the magneto-dielectric response and the magnetic field dependence of the dielectric constant~\cite{ikeda94b,subramanian06,serro08,wen09}; more than 10 percents of the dielectric constant are reduced by applying magnetic field within 0.2T near the magnetic ordering temperature.  
Recent neutron scattering experiments exhibit that diffraction intensity at the charge order superstructure at $(2/3 \ 2/3 \ 3.5)$ is strongly reduced by applying magnetic field~\cite{wen09}. 
Because of the large magneto-dielectric phenomena introduced above, this series of compounds is regarded as multiferroic materials. 

Charge and spin dynamics have also shown anomalous features. 
AC dielectric constant $\varepsilon(\omega) $ shows strong diffusive and dispersive nature~\cite{ikeda95,ikeda94}. 
Temperature dependence of $\varepsilon(\omega)$ in KHz and MHz regions in \lufo \ have a shallow dip structure around the charge order temperature. 
Observed characteristic step-like feature is attributed to a domain motion. 
The frequency dependence of the dielectric constant in low temperatures is fitted by the Debye model and data are scaled by the Cole-Cole plot. The temperature dependence obeys the activation type formula where the activation energy is about 0.3eV. The obtained relaxation frequency is about 1MHz which is comparable to a valence fluctuation frequency deduced from the M$\rm \ddot o$ssbauer experiments~\cite{tanaka89,tanaka84}. 
These results suggest the large charge fluctuation remains even below the charge ordering temperature, 
and is consistent with the recent optical experiments which will be introduced latter~\cite{xu08}.
It is demonstrated that charge order in the series of compound is controlled by applying voltage pulses and laser illumination~\cite{li09,li09b}, 
and is also sensitive to oxygen vacancies and ion substitutions~\cite{mori08,matsuo08,yoshii07}. 

\subsection{Mechanism of ferroelectric and magneto-electric phenomena}

As mentioned in the previous section, electronic structure in \rfo \  
is mainly responsible for the Fe $3d$ and O $2p$ electrons in the W-layers. Since equal amount of \fetwo and \fethr coexists, this material is considered as a frustrated system. 
Exotic dielectric and magneto-dielectric phenomena in \rfo \ have been examined from this point of view. 
In this section, theoretical examinations to resolve the anomalous dielectric and magneto-dielectric properties are reviewed. 

\begin{figure}[tb]
\begin{center}
\includegraphics[width=\columnwidth,clip]{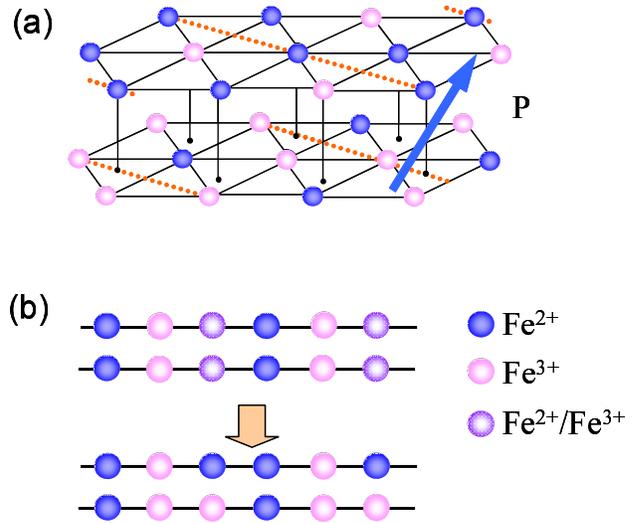}
\end{center}
\caption{ 
(a) Three-fold charge ordered structure in a W-layer.  
(b) Side views of the three-fold charge ordered state from [010] axis above and below three dimensional charge ordering temperature (after Ref.~\cite{yamada97}). 

}
\label{fig:wlayer}
\end{figure}
Microscopic mechanism of the dielectric properties in \rfo \ was studied by Yamada and coworkers~\cite{yamada97,ikedanature}. 
Schematic view of the proposed three-fold charge order model associated with the electric polarization is shown in Fig.~\ref{fig:wlayer}. 
In the low-temperature charge ordered phase, 
Fe ions are aligned as \fetwo-\fetwo-\fethr-\fetwo-\fetwo-\fethr in the upper plane and as \fetwo-\fethr-\fethr-\fetwo-\fethr-\fethr in the lower plane. 
As a result, charge distribution in a W-layer is imbalanced and the electric polarization appears. 
The authors also considered that, with increasing temperature, 
a partial disordered state without an electric polarization is realized. 
From this point of view, the incommensurate charge ordered state suggested by the satellite peak in the electron microscopy at $(1/3\pm 0.03 \ 1/3 \pm 0.03 \ l+1/2 )$ was examined. 

For the charge and spin structures in \rfo, 
the first principle electronic structure calculation based on the GGA+U scheme has been performed~\cite{xiang07,xiang09}. The authors claim that the three-fold charge order associated with the electric polarization in a W-layer is stabilized by lattice relaxation. 
From the calculated results of a small energy difference between this polar three-fold charge ordered structure and the other non-polar one, magneto-capacitance effect under magnetic field is discussed. 

Detailed microscopic electronic structure calculations based on the strong electron correlation has been performed by the present author's group~\cite{nagano07,nagano07b,naka08,nasu08,watanabe09}. 
These examinations are introduced in somewhat detail. 
Let us introduce, at the beginning, the electronic structure in \fetwo and \fethr ions. 
Iron ions are five fold coordinate, FeO$_5$, with $D_{3d}$ symmetry. 
The five $3d$ orbitals are split into the $d_{3z^2-r^2}$ orbital with the $A'$ symmetry,
the doubly degenerate $(-ad_{zx}+bd_{x^2-y^2}, ad_{yz}+bd_{xy})$ orbitals with the $E'$ symmetry, and 
the doubly degenerate $(bd_{zx}+ad_{x^2-y^2}, bd_{yz}-ad_{xy})$ orbitals with the $E''$ symmetry where 
$a$ and $b$ are the numerical coefficients. 
In the Madelung potential calculation, it is shown that the $E'$ orbitals have the lowest energy state, although the energy levels in the $E'$ and $E''$ orbitals are close with each other. 
For simplicity, it is assumed that the lowest energy state is the degenerate $(d_{xy}, d_{x^2-y^2})$ orbitals, 
and electrons in these orbitals are mobile and others are localized at Fe sites. 

Based on the electronic structure of the Fe ions, 
the tight-binding type Hamiltonian for the electronic structure in a W-layer is derived~\cite{nagano07,nagano07b,naka08}. 
The Hamiltonian is divided into the following terms,  
\begin{eqnarray}
{\cal H}={\cal H}_t+{\cal H}_V+{\cal H}_J . 
\label{eq:tjv}
\end{eqnarray}
The first term ${\cal H}_t$ is the electron hopping between the NN Fe ions, and the second term ${\cal H}_V$ is for the long-range Coulomb interactions. 
The third term ${\cal H}_J$ represents the superexchange interactions between the NN Fe ions, and 
is classified by valences of Fe ions, i.e. Fe$^{m+}$-Fe$^{n+}$ where $n$ and $m$ take $2$ and $3$. 
One of the representative term in ${\cal H}_J$ is given as 
\begin{eqnarray}
{\cal H}^{(22)} &=& J^{(22)} \sum_{\langle ij \rangle} 
\left ( {\bf I}_{i} \cdot {\bf I}_{j} + 6 \right) 
\nonumber \\
& \times& 
\left(\frac{1}{2} - 2 \tau_{i}^{ \eta} \tau_{j}^{\eta} \right) 
\left (\frac{1}{2} - Q_i \right) \left (\frac{1}{2} - Q_j \right) . 
\label{eq:hj1}
\end{eqnarray}
This term is for the superexchange interaction between NN Fe$^{2+}$ ions. 
A operator ${\bf I}_{i}$ represents spin in Fe$^{2+}$  with amplitudes of $2$, 
and $Q_i$ is the charge pseudo-spin operator which takes 1/2 for Fe$^{2+}$ and -1/2 for Fe$^{3+}$.  
The orbital degree of freedom in Fe$^{2+}$ ions is represented by the pseudo-spin operator $\tau_{i}^{\eta}$ where $\eta$ represents one of the three bond directions $(\alpha, \beta, \gamma)$ connecting sites $i$ and $j$. 
The prefactor $J^{(22)}$ is the exchange constant. 

\begin{figure}[tb]
\begin{center}
\includegraphics[width=\columnwidth,clip]{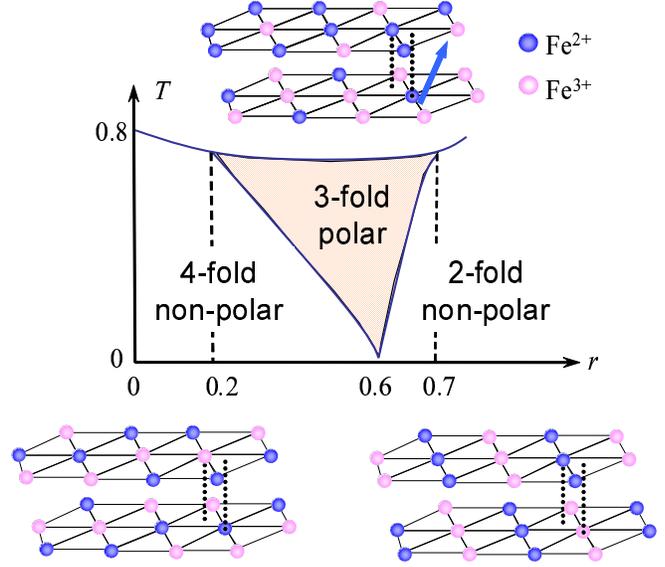}
\end{center}
\caption{ 
A mean-field phase diagram for charge order as function of temperature ($T$) and a ratio of the Coulomb interactions ($r$)~\cite{nagano07,naka08}. The fully frustrated point for a parameter set adopted in this figure corresponds to $r=0.6$.  
Schematic views of three-fold charge order, two-fold charge order and four-fold charge order are also shown. 
}
\label{fig:mf}
\end{figure}
Based on this Hamiltonian, the dielectric and magneto-dielectric phenomena have been investigated. 
At first, results obtained in the classical approaches are introduced~\cite{naka08}. 
The second and third terms in Eq.~(\ref{eq:tjv}) were analyzed by the mean-field approximation and the classical Monte-Carlo simulation at finite cluster. 
Figure~\ref{fig:mf} shows the mean-field phase diagram calculated in ${\cal H}_V$ as functions of temperature and a ratio of the Coulomb interactions $r=V_{cNNN}/V_{abNN}$ with the NN Coulomb interaction in $ab$ plane, $V_{abNN}$ and the second NN interaction along the $c$ axis, $V_{cNNN}$. 
This ratio $r$ measures magnitude of frustration, and $r=0.6$ corresponds to the fully frustrated point, $r_c$, for a parameter set adopted in Fig.~\ref{fig:mf}. 
It is shown that at $T=0$, the two-fold and four-fold charge ordered phases are stabilized. 
These phases do not show an electric polarization because of equal amount of \fetwo and \fethr in upper and lower plane in a W-layer. 
At $r=r_c$, the polar three-fold charge ordered state is degenerated with other two-fold and four-fold charge ordered states. 
In finite temperatures, the polar three-fold charge ordered state becomes stable due to thermal fluctuation effect. 
Large entropy gain in the three-fold charge ordered state is attributed to existence of the two sublattices; at sites in one of the two sublattices, the effective Coulomb interactions are canceled out and large charge fluctuation is easily induced.  

\begin{figure}[tb]
\begin{center}
\includegraphics[width=\columnwidth,clip]{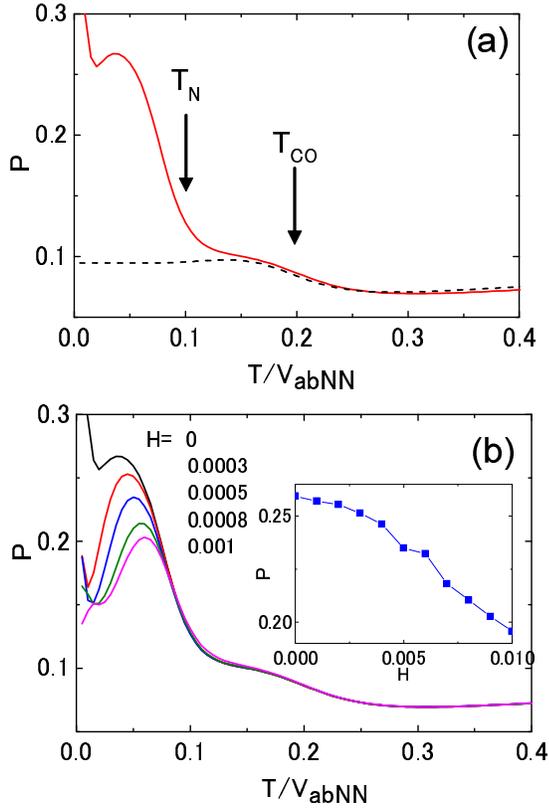}
\end{center}
\caption{ 
(a) Electric polarization correlation calculated in a model where the Coulomb interaction and the exchange interactions are taken into account~\cite{naka08}. Broken lines are the results obtained in a model where the Coulomb interactions are only considered. 
(b) Electric polarization correlation for various values of magnetic field $H$ as a unit of the electron transfer $t_{abNN}$~\cite{naka08}. 
Inset shows the magnetic field dependence of $P$. 
}
\label{fig:magdie}
\end{figure}
Effects of the spin degrees of freedom and charge-spin coupling were examined  
by analyzing the second and third terms in Eq.~(\ref{eq:tjv}), ${\cal H}_V+{\cal H}_J$~\cite{naka08}. 
Figure \ref{fig:magdie} shows the temperature dependence of the electric polarization correlation defined by $P=\sqrt{p^2}$ with 
$p=N^{-1}\sum_i(n_i^{U}-n_i^{L})$ where $n_i^{U(L)}$ is the electron number in the upper(lower) plane in a W-layer. 
A weak increasing in $P$ is seen at the charge ordering temperature. 
The polarization correlation shows a remarkable increase at the N$\rm \acute e$el temperature for a ferrimagnetic order and is saturated to the maximum value at zero temperature. The obtained ferrimagnetic structure is consistent with the neutron diffraction experiments. 
This large magneto-dielectric coupling is attributed to the spin-charge coupling in the superexchange interaction, ${\cal H}_J$, and the spin fluctuation due to the geometrical frustration as follows. 
The polar and non-polar phases with the three-fold charge order are competed with each other. 
Large spin entropy due to nearly partial spin order 
in the polar phase overcomes charge entropy gain in the non-polar state. 
In this point of view, this magneto-dielectric effect is caused by spin fluctuation based on frustration. 
It is demonstrated in Fig.~\ref{fig:magdie}(b) that the electric polarization is controlled by applying magnetic field. 

\begin{figure}[tb]
\begin{center}
\includegraphics[width=\columnwidth,clip]{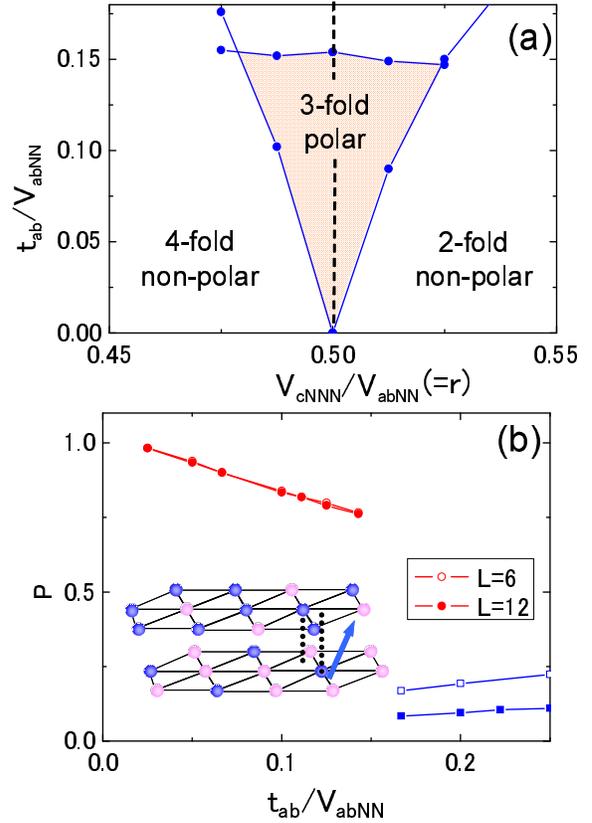}
\end{center}
\caption{
(a) Phase diagram for charge order as functions of the Coulomb interaction ratio and the electron transfer integral~\cite{watanabe09}. 
The fully frustrated point for a parameter set adopted in this figure corresponds to $r=0.5$.  
The transfer integral in the $ab$ plane $(t_{ab})$ and that in the $c$ axis $(t_{c})$ are assumed to be equal with each other.  (b) Polarization correlation at $V_{cNNN}/V_{abNN}=0.5$ corresponding to the broken line in (a)~\cite{watanabe09}. 
The system has $2L^2$ sites.}
\label{fig:quantum}
\end{figure}
%
Quantum fluctuation effects on the electronic ferroelectricity in a W-layer was examined in Ref.~\cite{watanabe09}. 
The first and second terms in Eq.~(\ref{eq:tjv}), ${\cal H}_V+{\cal H}_t$,  
were analyzed by the variational Monte-Carlo method. 
Figure \ref{fig:quantum}(a) shows the phase diagram at zero temperature as a function of the Coulomb interaction ratio $r( \equiv V_{cNNN}/V_{abNN})$ and the electron transfer integral. 
It is shown that the polar three-fold charge ordered phase is stabilized in a region of finite electron transfer. 
The polarization correlation, presented in Fig.~\ref{fig:quantum}(b), shows almost maximum value. 
Thus, the ferroelectric order is caused by a combination effect of the electron transfer between the triangle layers, i.e. quantum charge fluctuation, and the long-range Coulomb interaction. 
Two interactions induce asymmetric charge distribution between the two potential minima in a W-layer. 

\begin{figure}[tb]
\begin{center}
\includegraphics[width=\columnwidth,clip]{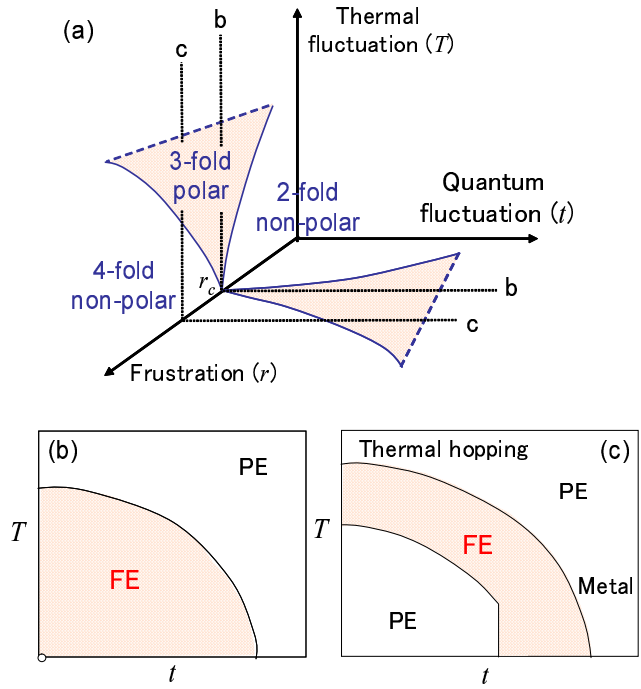}
\end{center}
\caption{ 
A schematic ferroelectric phase diagram for a W-layer as functions of frustration parameter $r$, temperature $T$ and transfer integral $t$. 
Lower panels show two-dimensional phase diagrams in the $T-t$ plane, when the frustration axis is cut at $b$ and $c$. 
}
\label{fig:fluctuation}
\end{figure}
A whole picture of this exotic dielectric property in a W-layer is summarized in Fig.~\ref{fig:fluctuation}. 
This is a schematic dielectric phase diagram as functions of the frustration parameter corresponding to $r$, temperature $T$ implying thermal charge fluctuation, and the electron transfer integral $t$ implying quantum charge fluctuation. 
This phase diagram is expected from the model where only the charge degree of freedom is taken into account, that is, ${\cal H}_t+{\cal H}_V$. 
At $r=r_c$, $T=0$ and $t=0$, a number of charge ordered states with and without electric polarization are largely degenerate. By including small values of thermal and quantum fluctuations at this point, the polar three-fold charge ordered state is stabilized. 
With further increasing the fluctuations, the ferroelectric order is monotonically weakened and finally the long range order disappears. 
Apart from the fully frustrated point of $r=r_c$ [see Fig.~\ref{fig:fluctuation}(c)], 
there is a paraelectric phase in low $T$ and small $t$ region. 
When thermal and quantum fluctuations increase, the paraelectric to ferroelectric transition occurs, and finally, thermal or quantum ferroelectric-paraelelctric transitions appear. 
These roles of thermal/quantum fluctuations on ferroelectricity are highly in contrast to these in usual displacive-type and the hydrogen-bond type ferroelectricities where fluctuations simply destroy ferroelectric order.  
Actually, remarkable dielectric fluctuations cause the dispersive and diffusive natures in the dielectric constant in \rfo \ as shown in the previous section. 
Recently, large charge fluctuation below the long-range charge ordering temperature was confirmed by the optical absorption spectra where temperature dependence of the absorption spectra due to \fetwo-\fethr are examined~\cite{xu08}. 
The experimental results support the above scenario for the exotic ferroelectricity in this material. 
It is worth noting that topological nature in the present phase diagram 
[Fig.~\ref{fig:fluctuation}(a)] is resemble to that for the spin driven type [Fig.~\ref{fig:mnphase}(b)]:  a V-shaped phase boundary around 
the critical frustration parameter and stabilization of a ferroelectric phase between 
the other types of long-range ordered phases. 
This common feature suggests that competition between the interactions due to geometrical frustration 
is of prime importance for the electronic ferroelectricity. 
When the spin degree of freedom, i.e. the exchange term ${\cal H}_J$, is taken into account, 
a ferroelectric phase appears in a finite region around $r_c$ even at $t=0$ and $T=0$. 
Although there is some uncertainly to determine a detailed value of the frustration parameter for real compounds, 
it is supposed that LuFe$_2$O$_4 $ is located around $r_c$, where the ferroelectric phase remains down to low temperatures. 
The frustration parameter for YFe$_2$O$_4$ is expected to be apart from  
$r_c$, since a sequential phase transition from the three-fold charge order to the other types of charge order is observed by lowering temperature~\cite{ikeda03}. 

As for the electronic structure beyond one W-layer, study for the interactions between the W-layers is introduced. 
Yamada and coworkers consider the interaction between the W-layers in the Ising-like model for the charge degree of freedom, and discuss the transition from two-dimensional charge order to the three-dimensional order with decreasing temperature~\cite{yamada00}. 
As for a possibility of the antiferroelectric structure suggested by the neutron diffraction experiments,  
an ab-initio band structure calculation based on the density-functional theory is applied to the two W-layers~\cite{xiang07}. The results show that energy for the antiferroelectric state, where the two electric moments in the two W-layers are antiparallel, is more stable than the ferroelectric state by a few meV per formula unit. 
The inter W-layer interactions were also studied by a phenomenological approach by Harris 
and coworkers~\cite{harris08}. 
Based on the Landau theory and symmetry considerations, the antiferroelectric transition associated with ferroelectric fluctuation above the transition temperature is realized in some parameter regions. 

Finally, the orbital state in \rfo \ is briefly mentioned. 
As introduced previously, the lowest orbital state in an isolated \fetwo ion is doubly degenerate under the crystalline field of the D$_{3d}$ symmetry and without the relativistic spin-orbit interaction. 
A possible orbital state in an ideal charge and spin ordered states in \rfo \ has been examined in Ref.~\cite{nasu08}.  
It is shown that the orbital state is able to be mapped on to a honeycomb lattice orbital model given by 
\begin{eqnarray}
{\cal H}=-J \sum_{i , \eta} 
\tau^{\eta}_i \tau^{\eta}_{i+{\bf e}_\eta}  ,  
 \label{eq:ham}
\end{eqnarray}
where $\eta=(\alpha, \beta, \gamma)$ represents the three kinds of bonds in a honeycomb lattice, and $\tau^\eta_i$ is the orbital pseudo-spin operator defined in the rotating frame for the $\eta$ bond. 
Numerical studies in this model by Monte-Carlo simulation and exact diagonalization method show that there is a large number of degeneracy in the classical ground state, and no conventional long-range orbital order occurs at finite temperature in the classical model, or at zero temperature in the quantum model. 
Recently, the orbital state in \lufo \ was examined by the resonant x-ray scattering experiments at Fe K-edge. The diffraction intensity at (1/3 \ 1/3 \ 1/2) does not show the azimuthal angle dependence that are expected in the conventional orbital long-range order~\cite{mulders09}. 
This suggests an orbital disordered state expected from the theoretical calculation. 
To clarify the orbital state in \rfo \ in more detail, further theoretical and experimental studies are required. 

\section{Summary and perspective}
We have reviewed progress of theoretical and experimental researches in electronic ferroelectricity from view point of geometrical frustration. 
Through the recent systematic and extensive studies in multiferroic materials, microscopic picture for spin driven ferroelectricity has been almost revealed. 
An electric polarization is induced by spin-lattice couplings by 
the magnetostriction effects through the antisymmetric and/or symmetric magnetic exchange interactions. 
Collinear and noncollinear spin textures with long periodicity resulted in geometrical frustration where spin interactions compete with each other. 
On the contrary, a whole picture of the charge driven ferroelectricity and roles of the frustration on electric polarization are still veiled. 
As an example of the charge driven ferroelectric materials, recent theoretical and experimental researches in 
the layered iron oxides are introduced. Combination effects of geometrical frustration and thermal/quantum charge fluctuation are of prime importance on the electric polarization. 
Further researches are required to confirm this picture.
For example, direct measurements of the charge fluctuation in wide momentum and frequency ranges by the inelastic x-ray scattering technique is useful. 
Time resolved techniques such as laser pump-probe experiments are valid to separate roles of charge and lattice degrees of freedom on ferroelectricity and to probe a unique charge dynamics.  
Systematic researches will provide a unified fashion of the charge driven ferroelectricity and new emergence of frustration.

\section*{Acknowledgment}
Author would like to thank A.~Nagano, M.~Naka, J.~Nasu, T.~Watanabe, H.~Takashima, and T.~Kimura for their collaborations. 
Author also would like to thank T.~Arima, Y.~Noda, N.~Ikeda, S.~Mori, Y.~Horibe, S.~Iwai, H.~Matsueda, and Y. Kanamori  for their valuable discussions. 
This work was supported by Grant-in-Aid for Scientific Research on Priority Areas ``Novel States of Matter Induced by Frustration'' (19052001), and ``High Field Spin Science in 100T'' (No.451) from the Ministry of Education, Culture, Sports, Science and Technology(MEXT), JSPS KAKENHI, and Grand challenges in next-generation integrated nanoscience. 
%


\end{document}